\documentstyle[twoside]{article}
\input epsf

\catcode`\@=11
\long\def\@makefntext#1{
\protect\noindent \hbox to 3.2pt {\hskip-.9pt
$^{{\eightrm\@thefnmark}}$\hfil}#1\hfill}       

\def\@makefnmark{\hbox to 0pt{$^{\@thefnmark}$\hss}}    

\def\ps@myheadings{\let\@mkboth\@gobbletwo
\def\@oddhead{\hbox{}
\rightmark\hfil\eightrm\thepage}
\def\@oddfoot{}\def\@evenhead{\eightrm\thepage\hfil
\leftmark\hbox{}}\def\@evenfoot{}
\def\sectionmark##1{}\def\subsectionmark##1{}}

\oddsidemargin=\evensidemargin
\newcounter{sectionc}\newcounter{subsectionc}\newcounter{subsubsectionc}
\renewcommand{\section}[1] {\vspace{12pt}\addtocounter{sectionc}{1}
\setcounter{subsectionc}{0}\setcounter{subsubsectionc}{0}\noindent
    {\tenbf\thesectionc. #1}\par\vspace{5pt}}
\renewcommand{\subsection}[1]
{\vspace{12pt}\addtocounter{subsectionc}{1}
    \setcounter{subsubsectionc}{0}\noindent
    {\bf\thesectionc.\thesubsectionc.
    {\kern1pt \bfit #1}}\par\vspace{5pt}}
\renewcommand{\subsubsection}[1]
{\vspace{12pt}\addtocounter{subsubsectionc}{1}
    \noindent{\tenrm\thesectionc.\thesubsectionc.\thesubsubsectionc.
    {\kern1pt \tenit #1}}\par\vspace{5pt}}
\newcommand{\nonumsection}[1] {\vspace{12pt}\noindent{\tenbf #1}
    \par\vspace{5pt}}

\newcounter{appendixc}
\newcounter{subappendixc}[appendixc]
\newcounter{subsubappendixc}[subappendixc]
\renewcommand{\thesubappendixc}{\Alph{appendixc}.\arabic{subappendixc}}
\renewcommand{\thesubsubappendixc}
    {\Alph{appendixc}.\arabic{subappendixc}.\arabic{subsubappendixc}}

\renewcommand{\appendix}[1] {\vspace{12pt}
        \refstepcounter{appendixc}
        \setcounter{figure}{0}
        \setcounter{table}{0}
        \setcounter{lemma}{0}
        \setcounter{theorem}{0}
        \setcounter{corollary}{0}
        \setcounter{definition}{0}
        \setcounter{equation}{0}
        \renewcommand{\thefigure}{\Alph{appendixc}.\arabic{figure}}
        \renewcommand{\thetable}{\Alph{appendixc}.\arabic{table}}
        \renewcommand{\theappendixc}{\Alph{appendixc}}
        \renewcommand{\thelemma}{\Alph{appendixc}.\arabic{lemma}}
        \renewcommand{\thetheorem}{\Alph{appendixc}.\arabic{theorem}}
        \renewcommand{\thedefinition}{\Alph{appendixc}.
        \arabic{definition}}
        \renewcommand{\thecorollary}{\Alph{appendixc}.
        \arabic{corollary}}
        \renewcommand{\theequation}{\Alph{appendixc}.\arabic{equation}}
        \noindent{\tenbf Appendix \theappendixc #1}\par\vspace{5pt}}
\newcommand{\subappendix}[1] {\vspace{12pt}
        \refstepcounter{subappendixc}
        \noindent{\bf Appendix \thesubappendixc. {\kern1pt \bfit #1}}
    \par\vspace{5pt}}
\newcommand{\subsubappendix}[1] {\vspace{12pt}
        \refstepcounter{subsubappendixc}
        \noindent{\rm Appendix \thesubsubappendixc.
        {\kern1pt \tenit #1}}
    \par\vspace{5pt}}

\newcommand{\textlineskip}{\baselineskip=13pt}
\newcommand{\smalllineskip}{\baselineskip=10pt}

\def\eightcirc{
\begin{picture}(0,0)
\put(4.4,1.8){\circle{6.5}}
\end{picture}}
\def\eightcopyright{\eightcirc\kern2.7pt\hbox{\eightrm c}}

\newcommand{\copyrightheading}[1]
    {\vspace*{-2.5cm}\smalllineskip{\flushleft
    {\footnotesize International Journal of Modern Physics B, #1}\\
    {\footnotesize $\eightcopyright$\, World Scientific Publishing
     Company}\\
     }}

\newcommand{\publisher}[2]{{\begin{center}\footnotesize\smalllineskip
    Received #1\\
    Revised #2
    \end{center}
    }}

\def\abstracts#1#2#3{{
    \centering{\begin{minipage}{4.5in}\baselineskip=10pt\footnotesize
    \parindent=0pt #1\par
    \parindent=15pt #2\par
    \parindent=15pt #3
    \end{minipage}}\par}}

\renewenvironment{thebibliography}[1]           
    {\frenchspacing
     \ninerm\baselineskip=11pt
     \begin{list}{\arabic{enumi}.}
    {\usecounter{enumi}\setlength{\parsep}{0pt}
     \setlength{\leftmargin 12.7pt}{\rightmargin 0pt} 
     \setlength{\itemsep}{0pt} \settowidth
    {\labelwidth}{#1.}\sloppy}}{\end{list}}

\newcounter{itemlistc}
\newcounter{romanlistc}
\newcounter{alphlistc}
\newcounter{arabiclistc}

\newcommand{\fcaption}[1]{
        \refstepcounter{figure}
        \setbox\@tempboxa = \hbox{\footnotesize Fig.~\thefigure. #1}
        \ifdim \wd\@tempboxa > 5in
           {\begin{center}
        \parbox{5in}{\footnotesize\smalllineskip Fig.~\thefigure. #1}
            \end{center}}
        \else
             {\begin{center}
             {\footnotesize Fig.~\thefigure. #1}
              \end{center}}
        \fi}

\newcommand{\tcaption}[1]{
        \refstepcounter{table}
        \setbox\@tempboxa = \hbox{\footnotesize Table~\thetable. #1}
        \ifdim \wd\@tempboxa > 5in
           {\begin{center}
        \parbox{5in}{\footnotesize\smalllineskip Table~\thetable. #1}
            \end{center}}
        \else
             {\begin{center}
             {\footnotesize Table~\thetable. #1}
              \end{center}}
        \fi}

\def\@citex[#1]#2{\if@filesw\immediate\write\@auxout
    {\string\citation{#2}}\fi
\def\@citea{}\@cite{\@for\@citeb:=#2\do
    {\@citea\def\@citea{,}\@ifundefined
    {b@\@citeb}{{\bf ?}\@warning
    {Citation `\@citeb' on page \thepage \space undefined}}
    {\csname b@\@citeb\endcsname}}}{#1}}

\newif\if@cghi
\def\cite{\@cghitrue\@ifnextchar [{\@tempswatrue
    \@citex}{\@tempswafalse\@citex[]}}
\def\citelow{\@cghifalse\@ifnextchar [{\@tempswatrue
    \@citex}{\@tempswafalse\@citex[]}}
\def\@cite#1#2{{$\null^{#1}$\if@tempswa\typeout
    {IJCGA warning: optional citation argument
    ignored: `#2'} \fi}}

\def\pmb#1{\setbox0=\hbox{#1}
    \kern-.025em\copy0\kern-\wd0
    \kern.05em\copy0\kern-\wd0
    \kern-.025em\raise.0433em\box0}

\def\fnt#1#2{\footnotetext{\kern-.3em
    {$^{\mbox{\scriptsize #1}}$}{#2}}}

\def\fpage#1{\begingroup
\voffset=.3in
\thispagestyle{empty}\begin{table}[b]\centerline{\footnotesize #1}
    \end{table}\endgroup}

\def\runninghead#1#2{\pagestyle{myheadings}
\markboth{{\protect\footnotesize\it{\quad #1}}\hfill}
{\hfill{\protect\footnotesize\it{#2\quad}}}}
\font\tenrm=cmr10
\font\tenit=cmti10
\font\tenbf=cmbx10
\font\bfit=cmbxti10 at 10pt
\font\ninerm=cmr9
\font\nineit=cmti9
\font\ninebf=cmbx9
\font\eightrm=cmr8

\def\qed{\hbox{${\vcenter{\vbox{            
   \hrule height 0.4pt\hbox{\vrule width 0.4pt height 6pt
   \kern5pt\vrule width 0.4pt}\hrule height 0.4pt}}}$}}

\def\bsc{{\sc a\kern-6.4pt\sc a\kern-6.4pt\sc a}}   
\def\bflatex{\bf L\kern-.30em\raise.3ex\hbox{\bsc}\kern-.14em
T\kern-.1667em\lower.7ex\hbox{E}\kern-.125em X}

\begin{document}

\runninghead{PrBa$_2$Cu$_3$O$_{7-y}$: Superconducting or Anomalously
Magnetic?}{PrBa$_2$Cu$_3$O$_{7-y}$: Superconducting or Anomalously
Magnetic?}

\normalsize\textlineskip
\thispagestyle{empty}
\setcounter{page}{1}

\copyrightheading{}         

\vspace*{0.88truein}

\fpage{1}
\centerline{\bf PrBa$_2$Cu$_3$O$_{7-y}$:}
\vspace*{0.035truein}
\centerline{\bf SUPERCONDUCTING OR ANOMALOUSLY MAGNETIC?}
\vspace*{0.37truein}
\centerline{\footnotesize V. N. NAROZHNYI\footnote{Corresponding author.
E-mail: narozh@ns.hppi.troitsk.ru}}
\vspace*{0.015truein}
\centerline{\footnotesize\it Institute for High Pressure Physics,
Russian Academy of Sciences}
\baselineskip=10pt
\centerline{\footnotesize\it Troitsk, Moscow Region, 142092, Russia}
\baselineskip=10pt

\centerline{\footnotesize\it and}

\centerline{\footnotesize\it Institut f\"ur Festk\"orper- und
Werkstofforschung}
\baselineskip=10pt
\centerline{\footnotesize\it Dresden e.V., Postfach 270016, D-01171
Dresden, Germany}
\vspace*{10pt}
\centerline{\footnotesize D. ECKERT,
K. A. NENKOV
, G. FUCHS and K.-H. M\"ULLER}
\vspace*{0.015truein}
\centerline{\footnotesize\it Institut f\"ur Festk\"orper- und
Werkstofforschung}
\baselineskip=10pt
\centerline{\footnotesize\it Dresden e.V., Postfach 270016, D-01171
Dresden, Germany}
\vspace*{10pt}
\centerline{\footnotesize T. G. UVAROVA}
\vspace*{0.015truein}
\centerline{\footnotesize\it Institute of Crystallography,
Russian Academy of Sciences}
\baselineskip=10pt
\centerline{\footnotesize\it Leninski pr. 59, Moscow, 117333, Russia}
\vspace*{0.225truein}
\publisher{(31 May 1999)}{(revised date)}
\vspace*{0.21truein}
\abstracts{In $\rm PrBa_2Cu_3O_{7-y}$ (Pr123) single crystals grown by
the flux method the kink in the magnetic susceptibility $\chi_{ab}(T)$,
connected with antiferromagnetic ordering of Pr, disappears after field
cooling (FC) in a field $H\parallel ab$-plane whereas the kink in
$\chi_c(T)$ remains unchanged after FC in $H\parallel c-$axis. This seems
to be connected with the coupling between the Pr and Cu(2) sublattices.
The Curie constant $C$ determined from the data reported for
superconducting Pr123 crystals grown by the traveling-solvent floating
zone (TSFZ) method (Zou {\it et al}, Phys. Rev. Lett., 80, 1074 (1998))
is about one half of that for our flux-grown non-superconducting
crystals. Thus, we propose that concentration of Pr in TSFZ crystals
seems to be about one half of the nominal concentration for Pr123.
Therefore, we propose that superconductivity in TSFZ samples is connected
most probably with the partial substitution of Pr by nonmagnetic Ba.}{}{}


\textlineskip           
\vspace*{12pt}          


\noindent
Recently Zou $\em et~al.$\cite{Zou_PRL98} reported the observation of
bulk superconductivity (SC) for $\rm PrBa_2Cu_3O_x$ (Pr123) single
crystals grown by the traveling-solvent floating zone (TSFZ) method.
(Earlier traces of superconductivity were found for Pr123 thin
films.\cite{Blacks_PRB96}) This result is in sharp contrast
to earlier reports, in which it was generally accepted, that Pr123 is the
only nonsuperconducting compound among the orthorhombic fully doped $R\rm
Ba_2Cu_3O_{7-y}$ ($R=$Y, rare earth) cuprates.\cite{Radousky}

In this work we have studied the magnetic properties of high quality
Al-free orthorhombic Pr123 single crystals grown in Pt crucibles by the
flux method. Atomic absorption spectroscopy analysis has shown that the
Pt contamination does not exeed 3$\cdot$10$^{-3}$~at.~\%. X-ray analysis
has revealed single phase twinned orthorhombic material with lattice
parameters $a$=3.868, $b$=3.911, and $c$=11.702~\AA.

\begin{figure}[htbp]
\vspace*{13pt}
\epsfysize=6.6cm
\centerline{\epsfbox{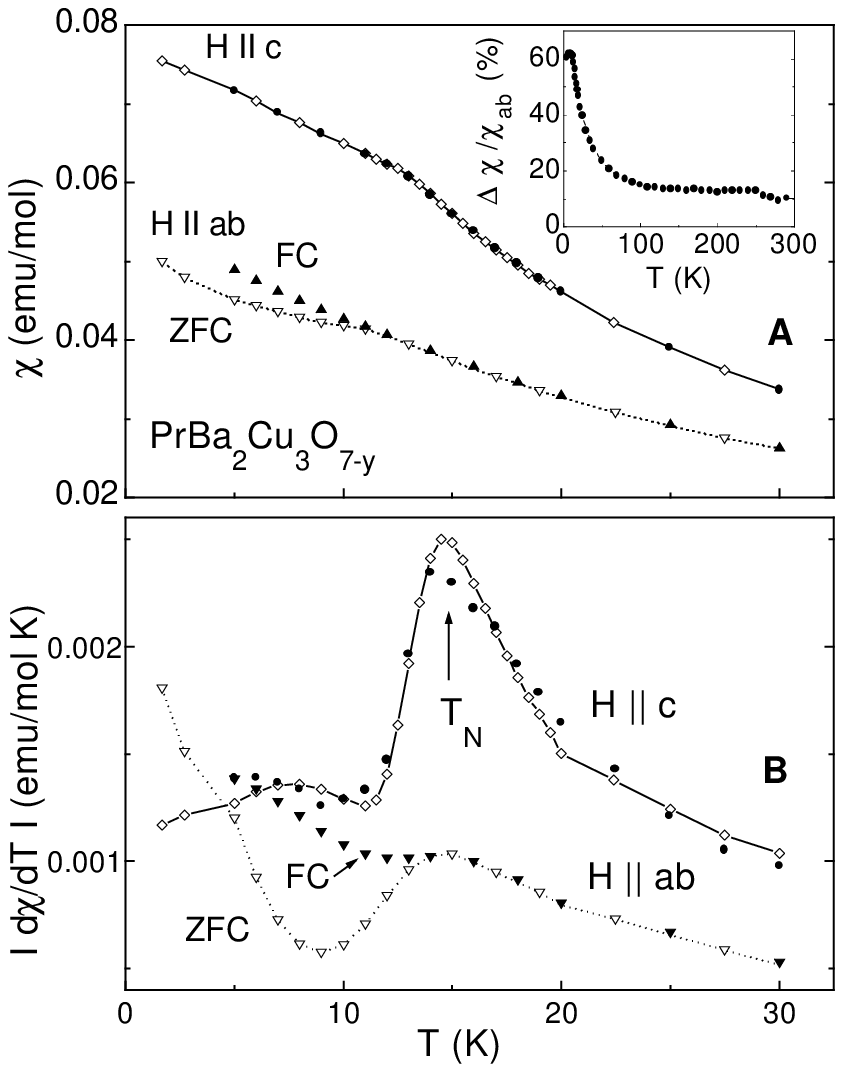}}
\vspace*{13pt}
\fcaption{(A) $\chi$ vs $T$ and (B) $|d \chi(T)/dT|$ vs $T$ for a Pr-123
single crystal. Lines connecting zero-field cooled (ZFC) data (open
symbols) are guides for the eye. Solid symbols represent the FC data. The
inset shows the anisotropy parameter $(\chi_c-\chi_{ab})/\chi_{ab}$ vs
$T$.}
\label{fig1}
\end{figure}

The anisotropy of the magnetic susceptibility $\chi$ is clearly seen in
Fig.\ \ref{fig1}; $\Delta\chi/\chi_{ab}=(\chi_c-\chi_{ab})/\chi_{ab}$ is
$\approx$10\% at $T$=300~K and increases with decreasing $T$ to
$\approx$60\% at $T$=15~K (see the inset of Fig.\ \ref{fig1}A). This
value is considerably larger than reported for crystals grown in alumina
crucibles\cite{Uma_PRB96} ($\approx$10\% at $T$=5~K) and is
close to the value reported recently for high quality Pr123 crystals
grown in BaZrO$_3$ crucibles\cite{Uma_JPh98}.

We have unexpectedly discovered that the kink in $\chi_{ab}(T)$
disappears after field cooling (FC) in a field $H\parallel ab$-plane,
whereas the kink in $\chi_c(T)$ remains unchanged after FC in $H\parallel
c$-axis. A possible explanation is connected with coupling of Pr and Cu
sublattices. (A theory of magnetic ordering in so called
exchange-frustrated antiferromagnets with two spin subsystems interacting
only by the anisotropic pseudo-dipole interaction has been recently
proposed by S.V. Maleev.\cite{Maleev_JETPL}) Pr ordering is accompanied
by a counter-rotation of the ordered Cu moments in bilayer with
establishing of a non-collinear Cu ordering below
$T_N$~\cite{Boothroyd_PRL97}. In our case freezing of the Cu magnetic
moments which lie in the $ab$-plane, by FC in $H\parallel$ $ab$-plane
hinders their re-orientation and, hence, the AFM ordering in the Pr
sublattice\cite{Nar_PhC99}. Our results may be considered as evidence for
the coupling of Pr and Cu(2) sublattices in Pr123.

\begin{figure}[htbp]
\epsfysize=4cm
\centerline{\epsfbox{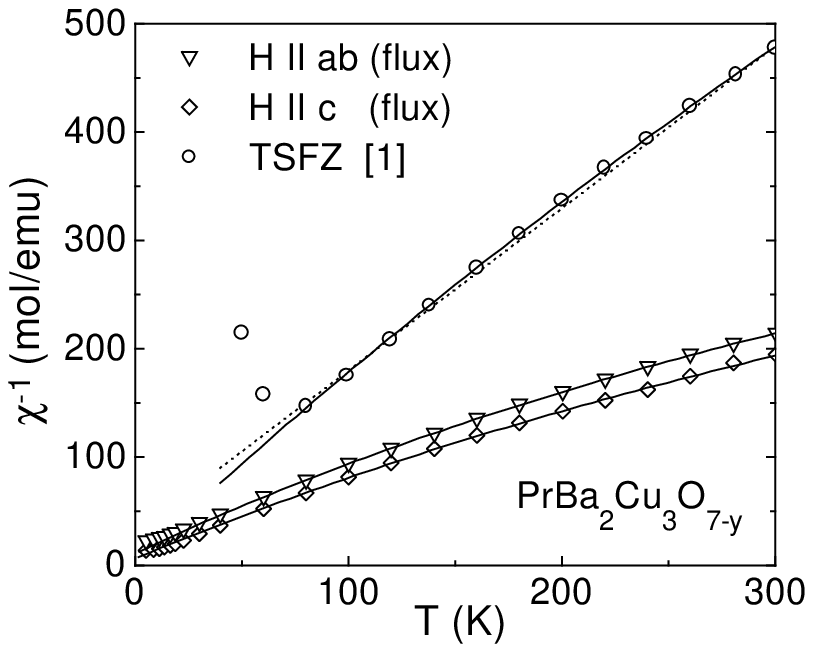}}
\vspace*{13pt}
\fcaption{$\chi^{-1}$ vs $T$ for flux grown
and TSFZ\protect \cite{Zou_PRL98}
Pr-123 single crystals. Solid lines -
fits to the Curie-Weiss law. Only some representative
points are shown. For details see text.}
\label{fig2}
\end{figure}

The values of $\mu_{eff}$=2.9~$\mu_B$ and 3.1~$\mu_B$ were obtained for
our crystal for $H||ab$-plane and $H||c$-axis, respectively. Zou $\em
et~al.$\cite{Zou_PRL98} report $\mu_{eff}$=2.92~$\mu_B$. At the same time
their $\chi(T)$ values are approximately two times lower than ours and
those reported by other groups\cite{Uma_PRB96,Uma_JPh98} for
non-superconducting samples, see Fig.\ \ref{fig2}. The value of
$\mu_{eff}$ was re-estimated\cite{Nar_PRL99} directly from the data of
Zou $\em et~al.$\cite{Zou_PRL98} The estimation gives
$\mu_{eff}$=2.09~$\mu_B$ and the Curie constant $C$=0.546~emu~K/mol. This
$C$ value for Zou's crystal is about one half of that for our flux
crystal (1.04~emu~K/mol and 1.19~emu~K/mol for $H||ab$ and $H||c$,
respectively). Based on this result, we propose that Pr occupies only
about one half of the $R$ sites (assuming for the TSFZ crystal nearly the
same Pr paramagnetic local moment as for the flux grown one). We propose
that the other half is occupied most probably by the nonmagnetic Ba. It
should be noted, that recently SC with $T_c \approx$~97~K was observed
for $\rm Pr_{0.5}Ca_{0.5}Ba_2Cu_3O_{7-y}$ bulk samples prepared under
high pressure\cite{Yao_PhC97}. Ba$^{2+}$ has a larger ionic radius than
Pr$^{3+}$, so the substitution of Ba on Pr site could give a natural
explanation not only for the superconductivity in the TSFZ Pr123 but also
for the elongation of distance between CuO$_2$ planes observed by Zou
{\it et al.}\cite{Zou_PRL98}.



\nonumsection{Acknowledgements}

\noindent
We thank S.-L. Drechsler and I. I. Mazin for useful discussions.
This work was supported by DFG grant MU1015/4-1 and RFBR grant
96-02-00046G

\nonumsection{References}

\end{document}